\documentstyle[12pt]{article}
\def\axy{A^{(X,Y)}}
\def\fxy{F^{(X,Y)}}

\def\b{\beta}

\def\d{\delta}
\def\e{\epsilon}
\def\t{\tau}

\def\s{\sigma}

\def\o{\omega}

\def\pa{\partial}

\def\O{\Omega}

\newskip\humongous \humongous=0pt plus 1000pt minus 1000pt
\def\caja{\mathsurround=0pt}
\def\eqalign#1{\,\vcenter{\openup1\jot \caja
	\ialign{\strut \hfil$\displaystyle{##}$&$
	\displaystyle{{}##}$\hfil\crcr#1\crcr}}\,}
\newif\ifdtup

\def\ref#1{$^{#1)}$}
\begin{document}
\begin{titlepage}
\begin{center}
January 1993            \hfill    BONN-HE-93/06
\vskip .4in

{\large \bf Geometry and Physics on $w_{\infty}$ Orbits.}
\footnote {This work was supported in part by the Deutsche
Forschungsgemeinschaft}

\vskip .2in
K. G. Selivanov
\footnote{on leave of absence from ITEP, Moscow }
\vskip .05in
{\em Physikalisches Institut der Universit\"at Bonn \\
Nu{\ss}allee 12, D-5300 Bonn 1 \\
Germany}
\end{center}

\vskip .05in

\begin{abstract}
We apply the coadjoint orbit technique to the group of area preserving
diffeomorphisms (APD) of a 2D manifold, particularly to the APD
of the semi-infinite cylinder which is identified with $w_{\infty}$.
The geometrical action obtained is relevant to both $w$ gravity and
2D turbulence. For the latter we describe the hamiltonian, which
appears to be given by the Schwinger mass term, and discuss some possible
developments within our approach. Next we show that the set of highest
weight orbits of $w_{\infty}$ splits into subsets, each of
which consists of highest weight orbits of $\bar{w}_N$ for a given N. We
specify the general APD geometric action to an orbit of $\bar{w}_N$ and
describe an appropriate set of observables, thus getting an action
and observables for $\bar{w}_N$ gravity. We compute also the Ricci form on the
$\bar{w}_N$ orbits, what gives us the critical central charge of the
$\bar{w}_N$ string,
which appears to be the same as the one of the $W_N$ string.
\end{abstract}

\end{titlepage}

\newpage
\renewcommand{\thepage}{\arabic{page}}
\setcounter{page}{1}
\setcounter{footnote}{1}

\section{Introduction}

  $w_{\infty}$ - or, more generally, the algebra of area
preserving diffeomorphisms (APD) of a 2D manifold
- has appeared in mathematical
physics in very different problems. First, as it was explained by
V.I. Arnold [Arn], motion of an inviscid incompressible fluid (in
a space of any dimension) can be viewed as a hamiltonian flow on a
coadjoint orbit of APD. Then, $w_{\infty}$ appeared to be an
$N {\rightarrow}{\infty}$ limit
of Zamolodchikov's $W_N$ algebra [Ba]. Another remarkable appearence
of $w_{\infty}$ - as a symmetry of the $ c=1 $ string - was discovered in
[KlePo]
and [Wit]. In [Ko] APD was discussed in the context of 3D Cern-Simons theory.
After all, APD is apparently a symmetry of the Nambu-Goto string
action (and therefore, of the Polyakov's one), also it is a symmetry of the
2D QCD action, and we will not even try
to list all appearences of $w_{\infty}$ in the recent enormous developments of
the theory of integrable equations.

    In our consideration of $w_{\infty}$ here we adopt the coadjoint orbit
ideology.

    In section 2 we obtain the geometrical action (that is the kinetic term
corresponding to the Kirillov symplectic form on the coadjoint orbit) for
APD without specifying the underlying 2D manifold and also without
the central extension term. As known from [AlSha] and [Wie], the geometrical
actions for the group of the diffeomorphisms of the circle and for the loop
 group
(with the central extensions) are nothing but the Liouville and WZW actions
correspondingly. That is why it was interesting to look at the APD case.
The action (2.18)
(possibly with some contributions from lower dimentional
cycles and from the boundary of the manifold, see discussions of section 3 )
 has all the rights to serve
as an action for $w_{\infty}$ gravity in its 3D formulation.
Apart from the kinetic term we also discuss possible Hamiltonians,
particularly, the Hamiltonian for the motion of the ideal fluid. It was funny
to understand that the latter was given by the so called Schwinger mass
term ( we have to say in advance that the dual space to the Lie algebra
of APD is essentially the space of abelian gauge potentials). We present some
speculations about a possible utilization of the APD action in the recently
proposed theory of 2D turbulence [Po].

In section 3 we specify our 2D manifold to be the semi-infinite cylinder,
the area preserving vector fields on which form the standard $w_{\infty}$
(on the cylinder
it would be rather $w_{1+\infty}$ ). We then discuss the structure of the
coadjoint orbits. Actually, due to presence of
huge ideals in $w_{\infty}$, the set of orbits splits into subsets
consisting
of coadjoint orbits of $\bar{w}_N$'s (not to be confused with $w_N$'s and
$W_N$'s !). After
describing the
appropriate representatives of the orbits, we specify the geometric action
of the section 2 (including in addition a central extension term) thus
getting geometrical actions for $\bar{w}_N$. Apart from the action we
 describe the
set of observables. The action and the observables should be considered as the
ones for $\bar{w}_N$ gravity.

    In section 4 we discuss the possibility of geometrical formulation
of the critical $\bar{w}_N$ string, following the ideology of [BoRa] developed
for the case of the bosonic (``Virasoro") string. Computing the curvature
of the ghost bundle over the $\bar{w}_N$ orbit we get an (upper) critical
 central
charge which appeared to be the same as the one for the $W_N$ string [Pop].

In section 5 we give some prospects and open questions.

    We will not try to make our definitions and arguments rigorous; our
consideration will rather be of a heuristic character.

\section{Geometrical action and Hamiltonians}

The  Kirillov bracket on a coadjoint orbits as well as the entire coadjoint
orbit machinery is present now in physicist's folklore so we shall be
short in general definitions following basically notations of [AlSha].

    To describe the Lie algebra of APD of the manifold $M$, pick a measure
${\omega}$
on $M$ - which in 2D case we are concerned with here is also a symplectic form.
The element of LieAPD is a vector field ${\epsilon}$ on $M$ preserving ${\o}$,
that is $ {\cal L}_{\epsilon}{ \omega} = 0 $
with ${\cal L}_{\e}$ being a Lie derivative,
which implies that the one-form $ i_{\epsilon}{ \omega} $ is closed,
$$ d {\,} i_{\epsilon}{ \,}{ \omega} = O \eqno(2.1) $$
Equivalently one
can say
that ${\epsilon}$ is divergentless.

Further, for a given one-form $A$ on $M$ one can construct a linear function
on LieAPD as
$$ {\langle} A , {\epsilon}{ \rangle} ={ \int} i_{\epsilon} A {\omega}
 \eqno(2.2) $$
{}From (2.1) it follows that $A$ and $A+d{\varphi}$ define the same function
${\langle} A,{\;}{\rangle}$, so the dual space to LieAPD, (LieAPD)$^{\ast}$
is identified with
the space of one-forms on $M$ modulo exact forms (with the proper definition
of the space of the forms, particulary in the case of noncompact $M$).

To get an explicit parametrization of the group APD, pick (local)
coordinates (${\sigma},{\tau}$) of $M$ such that (locally)
$${\omega}=d{\tau} {\wedge} d{\sigma} \eqno(2.3)$$
Then an $(X,Y)$ element of APD acts as
$$({\sigma},{\tau}){\,}{ \rightarrow}{ \,} (X({\sigma},{\tau}),Y({\sigma},
{\tau})) \eqno(2.4)$$
with the obvious constraint
$${ \partial}(X,Y)/{\partial}({\sigma},{\tau})=
X,_{\sigma} Y,_{\tau} - X,_{\tau} Y,_{\sigma} = 1 \eqno(2.5)$$
$(x,y)$ will always stand for the inverse of $(X,Y)$.

The coadjoint action of APD can be defined by saying that $A$ transforms as
a one-form:
$$ \eqalign{(A_{\sigma}&({\sigma},{\tau}),A_{\tau}({\sigma},{\tau}))^{(X,Y)}\cr
&=(A_{X}(X,Y)X,_{\sigma}+A_{Y}(X,Y)Y,_{\sigma},A_{X}(X,Y)X,_{\tau}
+A_{Y}(X,Y)Y,_{\tau})\cr} \eqno(2.6) $$
The infinitisimal coadjoint action by a vector field ${\e}$ is, of course,
given by
$$ {\d}_{ \e} A ={\cal L}_{\e} A \eqno(2.7) $$

Some $(A_{\s},A_{\t})$ can be chosen as a representative of an orbit ${\cal
O}$.
Then (2.6) gives a convenient parametrization of  ${\cal O}$.

Now the Kirillov's form ${\Omega}$ at the point $(A_{\s},A_{\t})^{(X,Y)}$
of ${\cal O}$
on the tangent-to-${\cal O}$ vectors ${\d}_{{\e}_1}{\axy}$
and ${\d}_{{\e}_2}{ \axy}$
takes the value
$${ \Omega}_{\axy}({\d}_{{\e}_1}{\axy},{\d}_{{\e}_2}{\axy})=
{\langle}{\axy},[{{\e}_1},{{\e}_2}]{\rangle} \eqno(2.8)$$
Identifying ${\d}{\axy}$ in the explicit parametrization (2.6)
with (2.7), one easily
derives that the corresponding vector field ${\e}_1$ is given by
$$ {\e}_1^{\mu}{ \pa}_{\mu} = (Y,_{\t}{ \d}_1X - X,_{\t}{ \d}_1Y){\pa}_{\s} +
(-Y,_{\s}{\d}_1X + X,_{\s}{\d}_1Y){\pa}_{\t} \eqno(2.9) $$
To write out the symplectic form ${\Omega}$ (2.8) one needs to compute
the commutator of ${\e}_1$ (2.9) and ${\e}_2$ (similar formula). One could do
it straightforwardly however more useful is to note first that the commutator
of two APD vector fields (in 2D case) is a hamiltonian one and then to see
that the corresponding Hamiltonian $h_{[{\e}_1,{\e}_2]}$ reads
$$ h_{[{\e}_1,{\e}_2]}= {\d}_1X{\d}_2Y-{\d}_1Y{\d}_2X \eqno(2.10) $$
Indeed, from (2.1) follows that (locally)
$$ i_{{\e}_1}{\o}=dh_1 \eqno(2.11) $$
where, of course, the function $h_1$ is not defined $a{\,}priory$ on the
whole $M$. Hence the commutator is (locally) given by
$$ i_{[{\e}_1,{\e}_2]}\o=d{\{}h_1,h_2{\}} $$
where ${\{}{\,},{\,}{\}}$ is the Poisson bracket, corresponding to ${\o}$.
However, by definition,
$$ {\{}h_1,h_2{\}}=dh_2({\e}_1) \eqno(2.12) $$
the right hand side of which is defined globally on $M$ and can be also
computed as $i_{{\e}_2}{\o}({\e}_1)$, easily giving (2.10).

We are in a position now to write out the symplectic form:
$${ \Omega}_{\axy}({\d}_{{\e}_1}{\axy},{\d}_{{\e}_2}{\axy})=
{\int}{\fxy}({\d}_1X{\d}_2Y-{\d}_1Y{\d}_2X) \eqno (2.13) $$
where ${\fxy}$ is the ``curvature" for a ``gauge potential" ${\axy}$,
${\fxy}=d{\axy}$.

The corresponding geometrical action reads
$$ S={\int}dt{\o}(A_X(X,Y){\dot{X}}+A_Y(X,Y){\dot{Y}}) \eqno (2.14)$$
where the diffeomorphism $(X,Y)$ parametrizing a point of  ${\cal O}$ is
asumed to be time ($t$, not ${\t}$!) -dependent and ${\dot{X}}$ (${\dot{Y}}$)
is the time-derivative of $X$ ($Y$).

Let us compute the variation of $S$ (2.14) - to observe the correspondence to
${\Omega}$ (2.13) and for future use. Straightforwardly varying one gets
$$ {\d}S= {\int}dt {\o}F_{XY}(X,Y)({\d}X{\dot{Y}}-{\d}Y{\dot{X}})\eqno (2.15)$$
Taking the vector field ${\e}$ defined as in (2.9) and the vector field
${\zeta}$ defined as in (2.9) with ${\dot{X}}$ (${\dot{Y}}$) instead of ${\d}X$
(${\d}Y$) (of course, ${\e}$ and ${\zeta}$ are divergent free, for variations
of $X$ and $Y$ as well as their time-dependence
are restricted by the constraint (2.5)) and using (2.10) one sees that ${\d}S$
 (2.15) rewrites as
$$ {\d}S={\int}dt{\Omega}_{\axy}({\d}_{\e}{\axy},{\d}_{\zeta}{\axy})
\eqno (2.16)$$
as it should according to the canons of hamiltonian mechanics
(apparently, ${\d}_{\zeta}{\axy}=d/dt{\,}{\axy}$).

It is useful to rewrite the action (2.14) in a little different form. Making
change of variables $({\s},{\t}){\rightarrow}(x({\s},{\t}),y({\s},{\t}))$
in the integration over $M$ in (2.14) (recall that $(x,y)$ is the inverse of
$(X,Y)$) one gets
$$ S={\int}dt{\o}(A_{\t}({\s},{\t})({\dot{x}}y,_{\s}-x,_{\s}{\dot{y}})+
A_{\s}({\s},{\t})(-{\dot{x}}y,_{\t}+x,_{\t}{\dot{y}}))   \eqno(2.17) $$
The action (2.17) can be rewritten in an even nicer form after elevating
the constraint (2.5)
(which is, of course, the same for both $(X,Y)$ and $(x,y)$)
to the action and combining $A$ with a lagrange multiplier $A_t$ into
a one-form ${\hat{A}}$ on the $3D$ manifold $R^1{\times}M$ ($S^1{\times}M$)
$$S={\int}{\,}{\hat{A}}{\wedge}{\o}-{\hat{A}}{\wedge}dx{\wedge}dy
\eqno(2.18)$$
We feel a need to recall that from the triple $(A_t,A_{\t},A_{\s})$
only $A_t$ should be integrated over in the path integral quantization
of the theory.

The equations of motion of a geometrical action say that a system
rests at one point of the phase space. If one wishes to have more interesting
dynamics - for example dynamics of the ideal fluid - then one needs to
include some hamiltonian which could $a{\,}priory$ be
an arbitrary function on a phase space. In the case at hand such functions are
given by arbitrary "gauge invariant"
functionals of ${\axy}$. One obvious
sample is the "curvature" ${\fxy}$. Consider a function $f$ defined by
$$ {\fxy}=f{\o} \eqno(2.19)$$
An arbitrary functional of $f$, in particular, any power of $f$ also gives a
function on ${\cal O}$. Taking a local function of $f$ and integrating it
over $M$ with the measure ${\o}$, one obviously gets an invariant of the
coadjoint action of APD and hence an invariant of any hamiltonian flow on
${\cal O}$.
However one should keep in mind that on some orbits powers of $f$ are
ill-defined (before any quantization!), see section 3.

To describe the ideal fluid Hamiltonian we for simplicity take $M$ to be
a disc $D$ and in some parametrization $({\s},{\t})$ (consistent with  (2.2))
 pick the metric
$$ g_{\mu,\nu}dz^{\mu}{\otimes}dz^{\nu}=d{\s}{\otimes}d{\s}+
d{\t}{\otimes}d{\t} $$
The metric allows us to construct another function the ${\cal O}$,
$${\cal H}={1 \over 2}{\langle}{\axy},{\rho}_{\axy}{\rangle} \eqno(2.20)$$
where ${\rho}_{A}$ is a divergenless vector field given by
$$ {\rho}_{A}^{\mu}{\pa}_{\mu}=
(g^{{\mu},{\nu}}-{{\pa}^{\mu}{\pa}^{\nu} \over {\pa}^2})
A_{\nu}{\pa}_{\mu} \eqno(2.21) $$
with ${\pa}^2$ being laplacian.

 ${\cal H}$ (2.20) is the sought hamiltonian.
To see this, notice that the variation of the canonical action
$S-{\int}dt{\cal H}$ is equal to
$$ {\int}dt({\Omega}_{\axy}({\d}_{\e}{\axy},d/dt{\,}{\axy})-
{\langle}{\d}_{\e}{\axy},{\rho}_{\axy}{\rangle}) \eqno(2.22) $$
(we have utilized (2.16)).  (2.22) is equivalent to the statement that
$$d/dt{\,}{\fxy}+ {\d}_{{\rho}_{\axy}}{\,}{\fxy}=0$$
what, taking into account (2.7),(2.19) and (2.21), gives the equation of
of 2D flow of the ideal fluid ( with $f$ from (2.19) being nothing
but the vorticity).

The Hamiltonian (2.20) might have something to do with the recently
proposed theory of 2D turbulence [Po]. Canonically, the time-independent
distribution is given by
$\exp(-{\b}{\cal H})$  while the measure of integration (to compute
correlators) is defined from the symplectic form ${\Omega}$ (2.13).
Naively, there is no place for turbulence. For example, enstrophy,
being one of the APD invariants discussed above, can be just  moved
away from any correlator. However, the necessity of the ultraviolet
cut off in the integration over ${\cal O}$ (UV-cut off in a "physical"
fluid is provided by viscosity, as argued in [Po]; note also that on some
orbits of $w_{\infty}$ the Hamiltonian has to be regularized even in classical
case, see next section) changes the
situation. The flows are not hamiltonian any more and hence the orbits (and
their
invariants!) are not invariant. In this case something has to be changed
in the original definition of the correlators because the regularized
system, instead
of walking along the phase space ${\cal O}$, moves somehow transversally to it.
Polyakov points out in his paper an heuristic similarity of the estrophy flux
in turbulence theory with the axial anomaly in field theory.
In our approach there is a very natural place for this analogy.
A coadjoint orbit can be viewed as a factor of the group by an isotropy
subgroup, in group variables the latter appearing as a gauge symmetry
of the action. Moving transversally to the orbit looks very much like an
anomaly in this gauge symmetry.
We will return to this problem somewhere else.

\section{Action and Observables for $\bar{w}_N$ case}

The standard definition of ${w_{\infty}}$ [Ba] is given in terms
of generators
$$ w_m^j,{\,} m=0,{\pm}1,{\pm}2,...{\,} j=0,1,2,... \eqno(3.1) $$
with the Lie bracket
$$ [w_m^j,w_n^k]=(m(k+1)-n(j+1))w_{m+n}^{j+k} \eqno(3.2) $$
This algebra can be viewed as a Poisson algebra formed by
Hamiltonians
$$ h_m^j=-i{\t}^{j+1}\exp(im{\s}) \eqno(3.3) $$
with the Poisson bracket defined by the symplectic form (2.3) or,
equivalently, as the algebra of (analytical) hamiltonian vector fields,
related to (3.3) by the rule (2.11).
The underlying manifold is then identified naturally with the
semi-infinite cylinder $ {\t} \geq 0 $, $ {\s}+2{\pi}{\,} \sim{\,} {\s}$.
On the semi-infinite cylinder one need not make a difference between
area preserving and hamiltonian vector fields, furthermore on the semi-infinite
cylinder
$j=-1$ is forbidden by the requirement that vector fields from LieAPD
 must be tangent to the
boundary (${\t}=0$).

To describe (LieAPD)$^{\ast}$ take the dual basis ${f_m^j}$ defined by
$${\langle}f_m^j,w_n^k{\rangle}=i{\d}^{j,k}{\d}_{m+n,0} \eqno(3.4)$$
We will consider only orbits, corresponding to the highest weight
representations, that is orbits with representative $f$ of the type
$$f=\sum_j {\,}b^j f_0^j \eqno(3.5) $$
where $b^j$ are numbers.

The coadjoint action is defined as usual by
$$ {\langle}{\d}_{w_m^j}f,w_n^k{\rangle} +
{\langle}f, [w_m^j,w_n^k] {\rangle}=0 \eqno(3.6) $$

Now consider orbits whose representative (3.4) consists only of
a finite number of $f_0^j$, that is there exists such an $N$ that
$$b^N{\neq}0{\;} \mbox{and}{\;} b^j=0{\;} \mbox{for}{\;}j>N \eqno(3.7) $$
{}From the definitions (3.3)-(3.6) it is easy to see that an isotropy
subalgebra $I_N$ for an orbit of the
type (3.5), (3.7) is generated by
$$ w_0^j,{\,} j{\leq}N{\;}\mbox{and}{\;}w_m^j, {\,}j>N, \mbox{every}{\;}m
\eqno(3.8) $$
So the orbits (3.5), (3.7) are in fact orbits of $\bar{w}_{N+2}$, particularly
at
$N=0$ - Virasoro orbits. One can say that these orbits posess a hidden
$w_{\infty}$ symmetry but it is very well hidden (the algebra of constraints
forms an ideal in the algebra of observables).
Note that for the algebra (3.1), (3.2) with the central extension
 (which could be nontrivial only in the Virasoro subalgebra [Ba]
and is given by Gel'fand-Fuks cocycle) there exists a richer
zoo the types of isotropy subalgebras being in direct correspondence
to ones of the Virasoro coadjoint orbits discussed e.g. in [Wit2].

In the more explicit notations of section 2, the basic element $f_m^j$ can be
represented as $f_m^j({\s},{\t})$ (the function $f({\s},{\t})$ from (2.19))
$$f_m^j({\s},{\t})=(-1)^j {\pa}_{\t}^{j+1}{\d}({\t})\exp(im{\s}) \eqno(3.9) $$

Specifying $M$ of section 2 and substituting representative $A$ corresponding
to $f$ (3.5) with $f_m^j$ from (3.9) into the action (2.17) one gets
$$S=-\sum_j b_j \sum_{k=0}^{j+1} \left({j+1}\atop k\right){\int}
dtd{\s}y_k \dot{x}_{j+1-k} \eqno(3.10)$$
where $y_k$  ($x_x$) stands for $d^k/d{\t}^k{\,}y_{/{\t}=0}$
($d^k/d{\t}^k{\,}x_{/{\t}=0}$ ). The derivatives of $x$ and $y$
are not independent, the derivatives of $y$ being expressed in terms of
$x$ due to the constraint (2.5) (henceforth we say (2.5) meaning its analog
for $(x,y)$)
and boundary condition
$$y({\s},0)=0 \eqno(3.11)$$
For example, restricting eq.(2.5) to the boundary ${\t}=0$ gives
$$ y_1={1 \over x_{0,{\s}}} \eqno(3.12) $$
Hitting (2.5) by ${\pa}/ {\pa}{\t}$ and putting ${\t}=0$ gives
$$ y_2=-{{\pa}_{\s}(x_1x_{0,{\s}}) \over (x_{0,{\s}})^3}  \eqno(3.13) $$
and so on.

Consider for example the case $N=1$ ($\bar{w}_3$ - case). Plugging (3.12),
(3.13)
into (3.10) gives
$$ S={\int}dtd{\s}\left(-b^0{{\dot{x}}_0 \over x_{0,{\s}}}-
2b^1{{\dot{x}}_1 \over x_{0,{\s}}}+
b^1{{\dot{x}}_1(x_1x_{0,{\s}}),_{\s} \over (x_{0,{\s}})^3}\right) \eqno(3.14)$$

As we already said, $w_{\infty}$ admits a central extension [Ba] given by the
Gel'fand-Fuks cocycle defined on the boundary ${\t}=0$:
$${\alpha}({\e},{\zeta})={1 \over 48{\pi}}
{\int}d{\s}({\pa}^3_{\s}{\e}{\zeta}-
{\e}{\pa}^3_{\s}{\zeta})_{{\t}=0} \eqno(3.15) $$
where ${\e}$ and ${\zeta}$ are vector fields from LieAPD (recall that
by definition they are tangent to the boundary).

Switching the central extension results in c-term contributions to the
symplectic form ${\Omega}$ (2.13) and to the action (2.17) (or (3.10)), the
contributions to the action being the 2D-gravity Wess-Zumino term [Po2]:
$$ {\Delta}S=-{c \over 24}{\pi}{\,}{\int}dtd{\s}
\left({{\dot{x}}_{0,{\s}}x_{0,{\s}{\s}} \over (x_{0,{\s}})^2}-
{{\dot{x}}_0(x_{0,{\s}{\s}})^2 \over (x_{0,{\s}})^3} \right) \eqno(3.16) $$
where c is dual to the central element of the Lie algebra. Due to the fact
that switching the central extension affects only the Virasoro subalgebra,
all the formulae on the way to (3.16) are parallel to those in [AlSha]
and it is not fun to reproduce them here.

Thus we claim that the action given by the sum of the terms (3.10) and (3.16)
is the action for $\bar{w}_{N+2}$ gravity ((3.14) and (3.16) for
$\bar{w}_3$-case).

The observables for $\bar{w}_{N+2}$ gravity in the present formalism read
$$ w_m^j={\int}f^{(X,Y)}h_m^j{\o}    \eqno(3.17) $$
with $h_m^j$ from (3.3) and $f^{(X,Y)}$ is $f$ from (3.5),(3.9) moved
by the element $(X,Y)$ of (centrally extended) APD:
$$ f^{(X,Y)}({\s},{\t})=f(X({\s},{\t}),Y({\s},{\t}))-
c/24{\pi}{\;}{\pa}_{\t}{\d}({\t}){\,}s(X_0,{\s})
\eqno(3.18) $$
where ${\d}({\t})$ is the Dirac ${\d}$-function and $s(X,{\s})$ is
the Schwartzian derivative,
$$s(X,{\s})={{\pa}_{\s}^3X \over {\pa}_{\s}X}-
{3 \over 2} ({{\pa}_{\s}^2X \over {\pa}_{\s}X})^2 $$
Making in (3.17) a change of variables
$({\s},{\t}){\rightarrow}(x({\s},{\t}),y({\s},{\t}))$ one gets
$$\eqalign{w_m^j=i\sum_k{\,}b^k{\int}d{\s}&
{\pa}_{\t}^{k+1}(y^{j+1}\exp(imx))_{/{\t}=0}\cr & +
i{c \over 24{\pi}}{\d}^{j,0}{\int}d{\s}{s(x_0,{\s}) \over
x_{0,{\s}}}\exp(imx_0)
\cr} \eqno(3.19) $$
Note that due to the boundary condition (3.11) $w_m^j$ automatically
equals zero for $j>N$.

Quantization of the theory thus obtained will for sure be considered somewhere
else.

\section{Ricci Form on $\bar{w}_N$ Orbits}

 In this section we adopt the ideology developed in [BoRa] for the
case of the bosonic (``Virasoro") string. They suggested to consider
its Hilbert space as some holomorphic homogeneous bundle over general
coadjoint Virasoro orbit (an orbit of the type $diffS^1/S^1$ which proved
to be a K\"ahler manifold). The bundle is proposed to be a tensor product
of a matter (bosonic) bundle and of a ghost vacuum bundle over $diffS^1/S^1$,
the former being tuned in such a way that its curvature cancels the
curvature of the latter. As explained in [BoRa] the curvature of the ghost
bundle is given by the Ricci form of the base K\"ahler manifold ($diffS^1/S^1$
in their case). They computed the Ricci form of $diffS^1/S^1$ and it appeared
to be equal
$$Ric(L_m,L_n)=(-{26 \over 12}m^3+{1 \over 6}m){\d}_{m+n,0} \eqno(4.1) $$
(the generators of the Virasoro algebra $L_m$
are viewed in (4.1) as vector fields
on the coadjoint orbit and (4.1) gives a value of the Ricci form on them).
The number 26 in (4.1) gives the value of the critical central charge
for the bosonic ($w_0$) string.

We transfer all the ideology and a lot of technical detailes from [BoRa]
to $w_{\infty}$-case ( the technique of Ricci form computation for an
infinite-dimentional homogeneous K\"ahler manifold has been developed
in [Fr]).

First, one easily sees that $\bar{w}_N$ orbit (3.5), (3.7) has a homogeneous
complex structure consistent with the symplectic form, thus being
a homogeneous K\"ahler manifold. Indeed, the complex structure $J$ on the
orbit ${\cal O}$ is built in a canonical way from the following
decomposition of the algebra $w_{\infty}$:
$$w_{\infty}=I_N{\oplus}w_N^+{\oplus}w_N^- \eqno(4.2) $$
where $I_N$ is the isotropy subalgebra (3.8), $w_N^+$ ( $w_N^-$ ) is
spanned by \linebreak $w_m^j, {\;} j \leq N, {\,} m>0 {\;}
 ( w_m^j, {\;} j \leq N, {\,} m<0 )$.

Define a vector field ${\zeta}_{w_l^k}$ on ${\cal O}$ in such a way that
its action on functions on ${\cal O}$ is just the coadjoint action. Then
on two such vectors at point $f$ the symplectic form ${\O}$ (2.8)
takes the value
$${\O}_f({\zeta}_{w_l^k},{\zeta}_{w_m^j})=
ib^{k+j}{\d}_{m+l,0}l(k+j+2) \eqno(4.3) $$
with $b^k$ in (4.3) and (3.5), (3.7) being the same. The form (4.3) is
apparently consistent with the complex structure (4.2), hence ${\cal O}$
is a K\"ahler manifold.

The K\"ahler metric $g^K$ on vectors ${\zeta}_{w_l^k}$ and
${\zeta}_{w_{-l}^j}$ at a point $f$ takes the value
$$g_f^K({\zeta}_{w_l^k},{\zeta}_{w_{-l}^j})=
{\O}_f({\zeta}_{w_l^k},J{\zeta}_{w_{-l}^j})=
b^{k+j}l(k+j+2) \eqno(4.4)$$
We will need some more general Hermitian metric on ${\cal O}$
$$g_f({\zeta}_{w_l^k},{\zeta}_{w_{-l}^j})=g_l^{k+l} \eqno(4.5) $$
with $g_l^k=0$ at $k>N$.
Consider now the metric-compatible-(with the metric (4.5))-Hermitian
connection ${\nabla}_{{\zeta}_{w_l^k}}$ on ${\cal O}$ and define the
operator ${\phi}$ on $T{\cal O}_f$ as
$$ {\phi}_{{\zeta}_{w_l^k}}V = {\nabla}_{{\zeta}_{w_l^k}}V -
{\cal L}_{{\zeta}_{w_l^k}}V \eqno(4.6) $$
Obviously, ${\phi}$ is an ultralocal (tensorial) operator preserving
the complex structure and the metric. Due to these properties it is
easily computable in general [Fr], [BoRa], and in our case it reads
(one needs only the restriction of ${\phi}$ to $T{\cal O}_f^+$, so in the
formulae below $j \leq N,{\,}n>0$)
$$\eqalign{&(i \leq N){\;} \cr
&{\phi}_{{\zeta}_{w_0^i}}{{\zeta}_{w_n^j}}
=(i+1)n{\zeta}_{w_n^{i+j}} \cr} \eqno(4.7a) $$
$$\eqalign{&i>N \cr
&{\;}{\phi}_{{\zeta}_{w_m^i}}{{\zeta}_{w_n^j}}
=(-(j+1)m+(i+1)n){{\zeta}_{w_{m+n}^{i+j}}} \cr} \eqno(4.7b)$$
$$\eqalign{&m>0,{\,}i \leq N \cr
&{\phi}_{{\zeta}_{w_{-m}^i}}{{\zeta}_{w_n^j}}=
((j+1)m+(i+1)n){{\zeta}_{w_{n-m}^{i+j}}}
{\theta}(N-i-j+1){\theta}(n-m) \cr} \eqno(4.7c)$$
$$\eqalign{&m>0,{\,}i \leq N \cr
&{\phi}_{{\zeta}_{w_m^i}}{{\zeta}_{w_n^j}}=
-\sum_{i,j}((j+1)m+(i+1)(p+m))
{\theta}(N-i-j+1)\cr
&(g_{p+m}^{-1})_{r,j}g_p^{i+j,k}
{{\zeta}_{w_{p+m}^r}} \cr} \eqno(4.7d) $$
where $(g_p^{-1})_{r,j}$ is the matrix inverse of $(g_p)^{r,j}$
and ${\theta}(j)$ is 1 when $j>0$ and 0 otherwise.

The curvature $R$ of the connection ${\nabla}_{\zeta}$ in terms of
${\phi}$ reads
$$R({{\zeta}_{w_m^i}},{{\zeta}_{w_n^j}}){{\zeta}_{w_p^k}}=
\left([{\phi}_{{\zeta}_{w_m^i}},{\phi}_{{\zeta}_{w_n^j}}]-
 {\phi}_{{\zeta}_{[w_m^i,w_n^j]}}\right){{\zeta}_{w_p^k}} \eqno(4.8) $$
$R$ (4.8) is considered as an operator on $T{\cal O}_f^+$ and the Ricci form
is, by definition, the trace of it. Thus, to compute the Ricci form we need
only the diagonal elements of $R$ and they are easily seen to be
$$\eqalign{&diag_p^kR({\zeta}_{w_m^i},{\zeta}_{w_{-m}^j}) \cr
&=-\sum_s((s+1)m+(i+1)p)((k+1)m+(j+1)p) \cr
&{\theta}(N-i-j+1){\theta}(p-m)
{\theta}(N-i-s+1)(g_p^{-1})_{k,s}g_{p-m}^{i+s,j+k} \cr
&+\sum_s((s+1)m+(i+1)(p+m))((k-j+1)m+(j+1)(p+m)) \cr
&{\theta}(N-i-s+1){\theta}(N-k+1)(g_{p+m}^{-1})_{k-j,s}g_p^{i+s,k} \cr
&-2mp \cr} \eqno(4.9) $$
To obtain the Ricci form one needs to sum over $k$ (up to $N$) and over
positive $p$. From the fact that $g_p^{k,s}$ is indeed $g_p^{k+s}$
it follows that the diagonal element (4.9) is nonzero only at $i=j=0$.
Hence the Ricci form is expressed as
$$ \eqalign{Ric&({\zeta}_{w_m^i},{\zeta}_{w_n^j})=
{\d}^{i+j,0}{\d}_{m+n,0}\sum_p \cr
&(-{\theta}(p-m)
\sum_{s,k}((k+1)m+p)((s+1)m+p)(g_p^{-1})_{k,s}g_{p-m}^{s,k}\cr
&+\sum_{s,k}((k+1)m+p+m)((s+1)+p+m)(g_{p+m}^{-1})_{k,s}g_p^{s,k} \cr
&-2mp(N+1)) \cr} \eqno(4.10) $$
Arbitrary $g_p^{k+s}$ can be expanded as
$$g_p^{k+s}=\sum_{q=0}^Na_{q,p}^N{\d}^{k+s,N-q} \eqno(4.11) $$
and plugging (4.11) into (4.10) gives
$$ \eqalign{Ric&({\zeta}_{w_m^i},{\zeta}_{w_n^j})=
{\d}^{i+j,0}{\d}_{m+n,0}\sum_p \cr
&( -{\theta}(p-m)
\sum_{k,s}((k+1)m+p)((s+1)m+p){a_{0,p-m}^N \over a_{0,p}^N} \cr
&+\sum_{k,s}((k+1)m+p+m)((s+1)m+p+m){a_{0,p}^N \over a_{0,p+m}^N} \cr
&-2mp(N+1) ) \cr} \eqno(4.12) $$
We come to a subtle point. The sum over $p$ in (4.12) is divergent
for the K\"ahler metric (4.4) (that is for $a_{0,p}=b^Np(N+2)$). The
Ricci form on the orbit ${\cal O}$ of $w_{\infty}$ is ill-defined. To define it
somehow, note that there exists a class of Hermitian metrics on ${\cal O}$
for which a sum over $p$ in (4.12) is absolutely convergent. Indeed, one sees
that
$$a_{0,p}=A_1^Np^{N+3}+A_2^Np^{N+1}+ ... \eqno(4.13) $$
($A_k^N$ are parameters) and arbitrary $a_{q,p}^N$ for $q>0$ define such
a class of metrics and with (4.13) one can proceed with the computation of
the Ricci form, getting finally
$$Ric({\zeta}_{w_m^0},{\zeta}_{w_{-m}^0})=
(-{26+4N^3+24N^2+46N \over 12}m^3 + {N+1 \over 6}m) \eqno(4.14) $$
Remarkably, the Ricci form is apparently independent on the parameters
$A_k^N$ from (4.13). To explain this somehow, note that on a finite-dimensional
manifold the cohomology class defined by the Ricci form is independent of the
metric and the proof can be transfered to the infinite-dimensional case
provided the Ricci form is well defined (the sum is convergent). The term
$m^3$ defines the nontrivial cohomology class on ${\cal O}$ and its
independence
on the metric is very saticfactory. We don't have any explanation why the $m^1$
term in (4.14) is metric independent.

Note that within the class (4.13) one can come as close as necessary to the
K\"ahler metric while keeping the Ricci form convergent. This and the
independence of the
Ricci form of the parameters $A_k^N$ is a justification of our regularization.

{}From (4.14) one can read off the critical central charge $c_{N+2}$ and the
 intercept
${\alpha}_0^{N+2}$ (for the Virasoro mode) for the $\bar{w}_{N+2}$ string:
$$c_{N+2}=26+4N^3+24N^2+46N,{\;}{\;}
{\alpha}_0^{N+2}=1+{1 \over 6}N^3 + N^2 + {11 \over 6}N \eqno(4.15)$$
Amusingly, they are the same as for $W_{N+2}$ string.

\section{Conclusion}

The appearence of the $\bar{w}_N$ algebra may be considered a little more
abstractly than in section 3. First, the space of vector fields on the
boundary ${\Gamma}$ of some 2D manifold M is easily seen to be isomorphic
to the factor of the algebra of APD vector fields on $M$ with respect to the
subalgebra of vector fields disappearing on ${\Gamma}$. The latter is
in fact an ideal, so that the factor has a structure of a Lie algebra and is
isomorphic to the algebra of vector fields on $S^1$. There are smaller
ideals in APD in which vector fields dissappear on ${\Gamma}$ with
their $N$-th derivatives. The factor of APD with respect to such $N$-th
ideal is isomorphic to the $\bar{w}_{N+2}$ algebra.

 Physically speaking, higher
spin fields are remnants of the extra-dimension. 2D gravity ($w_0$ gravity)
can be considered as having $w_{\infty}$ symmetry, the latter being very well
hidden in a sense that the constraints form an ideal in the algebra of
observables. Then the constants $b^j$
in the lagrangean (3.10) look like condensates, spontaneously breaking the very
well hidden
$w_{\infty}$ symmetry further and further.

The present analysis of $\bar{w}_N$ doesn't help a lot in the understanding
of $W_N$.
It would be interesting to understand what are the first principles ruling the
deformation of $\bar{w}_N$ gravity to the $W_N$ one. A natural guess is that
the
deformation follows from a quantization of the former, work on which is
in progress.

Also very interesting would be to understand how to get a standard
two dimensional formulation of $w_{\infty}$ gravity [BHPRSShSt]
from our 3D one (see action (2.17) or (2.18)).

Concerning 2D turbulence, our approach seems to be very promising.
The Schwinger mass term is intimately related to the axial anomaly which
is field theory analog of the turbulence fluxes [Po]. It suggests
to make a fermionization of the turbulence problem which is interesting
by itself as well as in view of the conformal perspective in 2D
turbulence theory. We will be back to all these stuffs somewhere else.

Acknowledgments.

I am grateful to my collegues from the University of Bonn for their
kind hospitality, especially to Niels Obers whose information store
and whose knowledge of computer tricks helped me a lot.

Note added in proof.

After the work was completed, A.S.Gorsky informed me that
some of the results of section 2 intersected with the ones of [Khe].
And a few days later R.Mkrtchyan came to the University of Bonn
with a talk [MM] devoted to the geometrical action for APD of 2D manifold
(see first part of the section 2).

\centerline{\underline{\bf References}}
\begin{enumerate}
\item[[AlSha]] A. Alekseev and S. Shatashvili, Path Integral Quantization
of the Coadjoint Orbits of the Virasoro Group and 2D Gravity,
Nuclear Phys. B323 (1989) 719-733
\item[[Arn]] V.I. Arnold, Mathematical Methods of Classical Mechanics,
(Springer-Verlag, New-York, 1978)
\item[[Ba]] I. Bakas, The large-$N$ limit of extended conformal systems,
Phys. Lett. 228B (1989) 57
\item[[BHPRSShSt]] E. Bergshoeff et al, $w_{\infty}$ gravity, Phys. Lett. 243B
(1990) 350,
Quantization deforms $w_{\infty}$
to $W_{\infty}$ gravity, Nuclear Physics B363 (1991) 163
\item[[BoRa]] M.J. Bowick and S.G. Rajeev, The Holomorphic Geeometry of
Closed Bosonic String Theory and $DiffS^1/S^1$,
Nuclear Physics B293 (1987) 348-384
\item[[Fr]] D.S. Freed, in Infinite dimensional groups with applications,
ed. V. Kac, (Springer, Berlin, 1985)
\item[[Khe]] B. Khesin, Zapiski Nauchnykh Seminarov LOMI, 1991
\item[[KlePo]] I. Klebanov and A.M. Polyakov, Mod. Phys. Lett., A6 (1991) 3273
\item[[Ko]] I.I. Kogan, Mod. Phys. Lett. A7 (1992) 3717
\item[[MM]] R. Manvelyan and R. Mkrtchyan, Geometrical action for $w_{\infty}$
algebra, to appear
\item[[Po]] A.M. Polyakov, The Theory of Turbulence in 2D,
Preprint PUPT-1369 (1992), Preprint PUTP-1341 (1992)
\item[[Po2]] A.M. Polyakov, Mod. Phys. Lett., A11 (1987) 893
\item[[Pop]] C.N. Pope, A Review of $W$ String, Preprint CTP TAMU-30/92 (1992)
\item[[Wie]] P.B. Wiegman, Multivalued functionals and geometric approach
to quantization of relativistic particles and strings, preprint MIT (1988)
\item[[Wit]] E. Witten, Nucl. Phys. B373 (1992) 187
\item[[Wit2]] E. Witten, Coadjoint Orbits of the Virasoro Group,
Comm. Math. Phys. 114, 1-53 (1988)
\end{enumerate}

\end{document}